\documentstyle[12pt]{article}
\catcode`\@=11

\def\@normalsize{\@setsize\normalsize{15pt}\xiipt\@xiipt
\abovedisplayskip 14pt plus3pt minus3pt%
\belowdisplayskip \abovedisplayskip
\abovedisplayshortskip  \z@ plus3pt%
\belowdisplayshortskip  7pt plus3.5pt minus0pt}
\def\small{\@setsize\small{13.6pt}\xipt\@xipt
\abovedisplayskip 13pt plus3pt minus3pt%
\belowdisplayskip \abovedisplayskip
\abovedisplayshortskip  \z@ plus3pt%
\belowdisplayshortskip  7pt plus3.5pt minus0pt
\def\@listi{\parsep 4.5pt plus 2pt minus 1pt
            \itemsep \parsep
            \topsep 9pt plus 3pt minus 3pt}}

\def\underline#1{\relax\ifmmode\@@underline#1\else
        $\@@underline{\hbox{#1}}$\relax\fi}
\@twosidetrue
\relax

\catcode`@=12

\catcode`\@=11

\def\section{\@startsection{section}{1}{\z@}{3.5ex plus 1ex minus
   .2ex}{2.3ex plus .2ex}{\large\bf}}

\def\ps@headings{\def\@oddfoot{}\def\@evenfoot{}
\def\@oddhead{\hbox{}\hfill
        \makebox[.5\textwidth]{\raggedright\ignorespaces --\thepage{}--
        \hfill }}
\def\@evenhead{\@oddhead}
\def\subsectionmark##1{\markboth{##1}{}}
}

\ps@headings

\catcode`\@=12

\relax

\def\figcap{\section*{Figure Captions\markboth
        {FIGURECAPTIONS}{FIGURECAPTIONS}}\list
        {Fig. \arabic{enumi}:\hfill}{\settowidth\labelwidth{Fig. 999:}
        \leftmargin\labelwidth
        \advance\leftmargin\labelsep\usecounter{enumi}}}
 \relax
\def\tablecap{\section*{Table Captions\markboth
        {TABLECAPTIONS}{TABLECAPTIONS}}\list
        {Table \arabic{enumi}:\hfill}{\settowidth\labelwidth{Table 999:}
        \leftmargin\labelwidth
        \advance\leftmargin\labelsep\usecounter{enumi}}}
 \relax
\def\reflist{\section*{References\markboth
        {REFLIST}{REFLIST}}\list
        {[\arabic{enumi}]\hfill}{\settowidth\labelwidth{[999]}
        \leftmargin\labelwidth
        \advance\leftmargin\labelsep\usecounter{enumi}}}
 \relax

\catcode`\@=11

\def\marginnote#1{}
\newcount\hour
\newcount\minute
\newtoks\amorpm
\hour=\time\divide\hour by60
\minute=\time{\multiply\hour by60 \global\advance\minute by-
\hour}
\edef\standardtime{{\ifnum\hour<12 \global\amorpm={am}%
    \else\global\amorpm={pm}\advance\hour by-12 \fi
    \ifnum\hour=0 \hour=12 \fi
    \number\hour:\ifnum\minute<100\fi\number\minute\the\amorpm}}
\edef\militarytime{\number\hour:\ifnum\minute<100\fi\number\minute}
\def\draftlabel#1{{\@bsphack\if@filesw {\let\thepage\relax
  \xdef\@gtempa{\write\@auxout{\string
    \newlabel{#1}{{\@currentlabel}{\thepage}}}}}\@gtempa
    \if@nobreak \ifvmode\nobreak\fi\fi\fi\@esphack}
     \gdef\@eqnlabel{#1}}
\def\@eqnlabel{}
\def\@vacuum{}
\def\draftmarginnote#1{\marginpar{\raggedright\scriptsize\tt#1}}
\def\draft{\oddsidemargin -.5truein
        \def\@oddfoot{\sl preliminary draft \hfil
        \rm\thepage\hfil\sl\today\quad\militarytime}
        \let\@evenfoot\@oddfoot \overfullrule 3pt
        \let\label=\draftlabel
        \let\marginnote=\draftmarginnote
   
\def\@eqnnum{(\theequation)\rlap{\kern\marginparsep\tt\@eqnlabel}%
\global\let\@eqnlabel\@vacuum}  }
\def\preprint{\twocolumn\sloppy\flushbottom\parindent 1em
        \leftmargini 2em\leftmarginv .5em\leftmarginvi .5em
        \oddsidemargin -.5in    \evensidemargin -.5in
        \columnsep 15mm \footheight 0pt
        \textwidth 250mmin      \topmargin  -.4in
        \headheight 12pt \topskip .4in
        \textheight 175mm
        \footskip 0pt
        
\def\@oddhead{\thepage\hfil\addtocounter{page}{1}\thepage}
        \let\@evenhead\@oddhead \def\@oddfoot{} \def\@evenfoot{} 
}
\def\titlepage{\@restonecolfalse\if@twocolumn\@restonecoltrue\onecolumn
     \else \newpage \fi \thispagestyle{empty}\c@page\z@
        \def\thefootnote{\fnsymbol{footnote}} }
\def\endtitlepage{\if@restonecol\twocolumn \else  \fi
        \def\thefootnote{\arabic{footnote}}
        \setcounter{footnote}{0}}  
\catcode`@=12
\relax

\def\ps@headings{\def\@oddfoot{}\def\@evenfoot{}
\def\@oddhead{\hbox{}\hfill
        \makebox[.5\textwidth]{\raggedright\ignorespaces --\thepage{}--
        \hfill }}
\def\@evenhead{\@oddhead}
\def\subsectionmark##1{\markboth{##1}{}}
}

\ps@headings

\relax

\def\firstpage#1#2#3#4#5#6{
\begin{document}
\begin{titlepage}
\nopagebreak
\title{\begin{flushright}
        \vspace*{-1.8in}
        {\normalsize CERN-TH/98-111}\\[-9mm]
        {\normalsize hep-th/9804006}\\[-9mm]
        {\normalsize April 1998}\\[14mm]
\end{flushright}
\vspace{2cm}
{#3}}
\author{\large #4 \\[0.0cm] #5}
\maketitle
\vskip 6mm
\nopagebreak 
\begin{abstract}
{\noindent #6}
\end{abstract}
\vfill
\begin{flushleft}
\rule{16.1cm}{0.2mm}\\[-3mm]
{\small$^\dag$ e-mail: Sergio.Ferrara, Alexandros.Kehagias, 
Herve.Partouche, Alberto.Zaffaroni@cern.ch} \\[-3mm]
\end{flushleft}
\thispagestyle{empty}
\end{titlepage}}

\def\simlt{\stackrel{<}{{}_\sim}}
\def\simgt{\stackrel{>}{{}_\sim}}
\newcommand{\dal}{\raisebox{0.085cm}
{\fbox{\rule{0cm}{0.07cm}\,}}}

\newcommand{\be}{\begin{eqnarray}}
\newcommand{\ee}{\end{eqnarray}}
\newcommand{\btau}{\bar{\tau}}
\newcommand{\p}{\partial}
\newcommand{\bp}{\bar{\partial}}
\newcommand{\cR}{{\cal R}}
\newcommand{\tR}{\tilde{R}}
\newcommand{\tcR}{\tilde{\cal R}}
\newcommand{\hR}{\hat{R}}
\newcommand{\hcR}{\hat{\cal R}}
\newcommand{\oE}{\stackrel{\circ}{E}}
\renewcommand{\p}{\partial}
\renewcommand{\bp}{\bar{\partial}}

\newcommand{\gsi}{\,\raisebox{-0.13cm}{$\stackrel{\textstyle
>}{\textstyle\sim}$}\,}
\newcommand{\lsi}{\,\raisebox{-0.13cm}{$\stackrel{\textstyle
<}{\textstyle\sim}$}\,}
\date{}
\firstpage{3118}{IC/95/34}
{\large $AdS_6$ {\Large I}NTERPRETATION OF 5D {\Large S}UPERCONFORMAL 
\\ {\Large F}IELD {\Large T}HEORIES
\phantom{X}\\\phantom{X}}
{S. Ferrara, A. Kehagias,  H. Partouche and A. Zaffaroni$^\dag$} 
{
\normalsize\sl Theory Division, CERN, 1211 Geneva 23, Switzerland
}
{We explore the connection of anti-de-Sitter supergravity in six dimensions,
based on the exceptional $F(4)$ superalgebra, and its boundary superconformal
singleton theory. The interpretation of these results in terms of a D4-D8
system and its near horizon geometry is suggested.
}

Recently, a close connection between superconformal field theories in $d$
 dimensions and anti-de-Sitter supergravities in $d+1$ dimensions has emerged \cite{kleb}-\cite{skend}. The original proposed duality between world-volume theories of $p$-branes 
systems
and their nearly horizon geometry in $AdS_{p+2}$ space-time \cite{malda} 
has further been interpreted as a correspondence between superconformal $p+1$
world-volume theories with $N$ (Poincar\'e) supersymmetries and $AdS_{p+2}$
supergravity with $2N$ supersymmetries. In addition,  the global symmetries 
of the 
former correspond to gauge symmetries of the latter \cite{fer,pol,witten}.

Apart from its physical interpretation, these recent developments have been
inherited by some peculiar properties of a class of unitary representations 
of the $O(p+1,2)$ conformal group, the so called singletons \cite{FlFr}, 
which have the properties that they allow propagation of massless degrees 
of freedom not on the $AdS_{p+2}$
bulk but rather on its boundary $\partial AdS_{p+2}=\tilde M_{p+1}$,
where $\tilde M_{p+1}$ is a certain completion of Minkowski space. Bulk degrees
of freedom turn out to be composite of singletons. This fact being closely 
related to the group property of decomposition of tensor products of 
singleton representations \cite{fer}. When this state of affairs is enlarged to incorporate supersymmetry, then a 
quite remarkable relation is discovered between BPS $p$-brane world-volume 
dynamics
and gauged superalgebras. 

The relevant superalgebras for $p=1,2,3$ and 5 are members of infinite 
sequences of superalgebras, usually denoted by $OSp(N/M)$ and $SU(N/M)$,
the orthosymplectic and unitary series \cite{nahm}.
However, in the particular case of a six-dimensional anti-de-Sitter space, 
there is an isolated superalgebra, $F(4)$, whose bosonic part is 
$SO(5,2)\times SU(2)$. It is the aim of the present work to study this case.

It turns out that the gauged $F(4)$ theory \cite{romans} has on its boundary a $N=2$ superconformal
field theory which is a superconformal fixed point of a five-dimensional Yang-Mills
theory. The relevant singleton representation of $F(4)\times G$ on the supergravity side is just a 
5d hypermultiplet in a symplectic representation of the flavour group $G$ at the fixed point of the Yang-Mills theory.

Five-dimensional superconformal field theories with some flavour symmetry 
group $G$ will correspond in the six dimensional bulk to the gauged $F(4)$ 
theory \cite{romans}, coupled to matter vector multiplets gauging the group 
$G$. \footnote{The Lagrangian for gauged supergravity coupled to matter 
vector multiplets has not yet been constructed.} These theories are based on a 
non-linear sigma-model \cite{romans,auria}
\be
R^+\times {SO(4,k)\over SO(4)\times SO(k)}\, ,
\label{manifold}\ee
where the diagonal $SU(2)$ in $SO(4)=SU(2)\times SU(2)$ is the gauged 
R-symmetry of the world-volume theory and we have additional gauge bosons corresponding
to the flavour symmetry $G$ ($k=\mbox{dim} G$) and a neutral $U(1)$ vector in 
the gravitational multiplet. The latter is absorbed by an antisymmetric 
tensor $B_{\mu\nu}$ to give a ``massive'' two-form \cite{romans}. 
The gauged bulk theory has a $G$-invariant anti-de-Sitter vacuum for a fixed value of
the {\it dilaton} and all matter scalars vanishing \cite{romans}. 
The gauged $F(4)$ theory, in the Poincar\'e limit, corresponds to the $d=6$, 
$(1,1)$ theory (with 16 supercharges).

One may ask why no superconformal extension exists for the $(2,2)$ theory,
contrary to the other cases in diverse dimensions. This can be understood from the fact that the $N=4$, 5d theory in Minkowski space-time  admits only 
vector multiplets and the latter are not conformal multiplets in five dimensions.
In other words, there is no candidate for a supersingleton multiplet for a 
hypothetical maximally extended theory. One may also ask what is the correspective
of this statement in terms of brane world-volume, in the spirit of 
\cite{malda}. $n$ D4-branes in type IIA realize on their world-volume a 
$U(n)$ Yang-Mills
theory which flows in the IR to a free (and not superconformal) theory of vector multiplets. The near horizon geometry of the D4-branes is not related to 
$AdS_6$. The
regime of validity of the supergravity solution (small curvature and small     
dilaton) in the large $n$ limit does not cover the IR region, as discussed in
\cite{IMSY}.

However, in the $N=2$ case the 5d hypermultiplets are conformal invariant and
there is a candidate supergravity multiplet for the corresponding $F(4)$
superalgebra. Non-trivial superconformal fixed points in $N=2$ Yang-Mills
theories in 5d were found in \cite{seib5}. A common characteristic of these
theories is that the global symmetries are enhanced at the fixed point. The R-symmetry 
for $N=2$ five-dimensional Yang-Mills theories is $SU(2)_R$ and we do not 
expect that it is enhanced at the superconformal point: it becomes the 
bosonic $SU(2)$ subgroup of the superalgebra $F(4)$. We will consider 
Yang-Mills theories with gauge group $USp(2n)$ with matter in the 
antisymmetric representations and $N_f$
fundamentals ($N_f\le 7$): the additional global symmetry is $SU(2)\times 
SO(2N_f)\times U(1)_I$, where $U(1)_I$ is associated with the current 
$*(F\wedge F)$, which exists and is conserved in  five dimensions. For 
generic values of the parameters, these theories are IR free. However, if we 
tune the parameters (in particular the {\it bare} coupling constant) in such 
a way that the effective coupling constant diverges at the origin of the 
Coulomb branch, we find non trivial
fixed points, where the global symmetry is enhanced to $SU(2)\times 
E_{N_f+1}$ ($E_5=Spin(10), E_4=SU(5), E_3=SU(3)\times SU(2), E_2=SU(2)\times 
U(1), E_1=SU(2)$) \cite{seib5}. Other fixed points with $U(1)$ or no global 
symmetry at all were constructed in \cite{ms}. The theory at the fixed point is a 
superconformal theory of interacting hypermultiplets,
with a global symmetry  $SU(2)\times E_{N_f+1}$. The global symmetry quantum 
numbers are carried  by instantons, which are particles in five dimensions 
and are the only states charged under $U(1)_I$ \cite{seib5}. Instantons can 
become
massless exactly when the coupling constant diverges. If we give mass to the 
adjoint we get another series of fixed points with $E_{N_f+1}$ global symmetry.

Evidence for non-trivial fixed points for other simple gauge groups coupled 
to matter
in various representations can be found in \cite{msi}. There is a reason
for having discussed the $USp(2n)$ case. It is the case 
which admits a brane realization in terms of D4 and D8-branes. The 5d fixed
points were indeed originally found by analyzing a D4-D8-brane configuration 
\cite{seib5}. We briefly review the construction. Consider type I$'$ on 
$S^1/{\bf Z}_2$. There are two orientifold planes and we will consider $2N_f$ D8 
branes coinciding with one of them. The $USp(2n)$ Yang-Mills theory, with 
exactly the matter content described above, is
obtained on the world-volume of $2n$ D4-branes living at the same orientifold.
The global symmetry of the D4-branes theory is $SU(2)\times SO(2N_f)\times 
U(1)_I$. The $SU(2)$ factor comes from the space-time Lorentz group; it is 
the less
relevant factor in our discussion, and disappears if we give a mass to the 
matter in the antisymmetric representation. $SO(2N_f)$ is the gauge symmetry 
of the $N_f$ D8-branes nine-dimensional world-volume theory; it is a subgroup 
of the $SO(32)$ type I gauge group. $U(1)_I$ corresponds to the $U(1)$ vector 
field of the
type IIA theory. 
The enhancement of global symmetry can be understood using the duality with
the $SO(32)$ heterotic string \cite{seib5}. 
In nine dimensions, T-duality connects the $SO(32)$ and $E_8\times E_8$ 
heterotic strings.
The type I$'$ backgrounds with enhanced symmetry correspond to the points
in the moduli space of the $SO(32)$ heterotic string on $S^1$ where there is 
an enhancement of symmetry to $E_8\times E_8$, or subgroups. The $U(1)_I$ 
charge
is the winding number of the dual heterotic string and the perturbative 
enhancement of space-time symmetry takes place at points in the moduli space 
where heterotic winding modes become massless. In the previous examples, 
$SO(2N_f)\times U(1)_I$ is enhanced
to $E_{N_f+1}$ . 

For our purpose the details of the type I$'$ theory are
irrelevant: we are describing a system of $2n$ D4-branes in the background
of $2N_f$ D8-branes at a nine-dimensional orientifold plane. The value of the 
dilaton at the orientifold diverges at the point in moduli space where we 
expect enhanced symmetry \cite{wp,seib5}; the heterotic winding modes are 
D0-branes in type I$'$ and
therefore become massless, providing the extra gauge bosons needed to fill the
adjoint of $E_{N_f+1}$ \cite{notes,lift}. The value of the dilaton at the 
orientifold is, for the five-dimensional theory on the D4-branes, the
effective coupling constant at the origin of the Coulomb branch, 
which is therefore tuned to infinity. The D0-branes are instantons for the 
D4-branes theory. This gives the picture  of a non-trivial superconformal 
fixed point, with relevant degrees of freedom corresponding to instantons, 
obtained by tuning to zero the inverse coupling
constant \cite{seib5}.

It is amusing to notice that the enhancement of the flavour symmetry $O(2N_f)\times O(2)\rightarrow E_{N_f+1}$ at the superconformal fixed point is the compact version of the enhancement of the T-S duality group to the U duality group
$O(10-d,10-d)\times O(1,1)\rightarrow E_{11-d(11-d)}$, obtained by replacing 
$N_f\rightarrow 10-d$ \cite{solvable}. 

The spinor representations of $SO(2N_f)$ ($SO(10-d,10-d)$), appearing in the decomposition of the adjoint of
$E_{N_f+1}$ ($E_{11-d(11-d)}$) and providing the missing vector bosons of the enhanced symmetry, can be obtained in the brane description by  quantizing
 the modes corresponding to the D0-D8 open strings \cite{notes,lift}.

Let us examine the implications of a $AdS_6$ supergravity description for the
superconformal fixed points with $E_{N_f+1}$ global symmetry. The D4-D8 system
has a global symmetry $SU(2)_R\times SU(2)\times E_{N_f+1}$. For simplicity,
we do not consider the $SU(2)$ factor in the following analysis: the corresponding superconformal theory is obtained as a limit of the same $USp(2n)$
Yang-Mills theory  with a mass term
for the multiplet in the antisymmetric representation. These fixed points will correspond to the gauged $F(4)$ supergravity \cite{romans} coupled to
matter vector multiplets in the adjoint of $E_{N_f+1}$. The superconformal
theory at the boundary is a theory of singleton hypermultiplets transforming
in a symplectic representation of $E_{N_f+1}$. The massless bulk 
supermultiplets can be identified with
bilinear composite operators on the boundary, corresponding to the
supermultiplets of global currents \cite{fer}. The multiplets are classified 
according to the maximal subgroup $USp(4)\times O(2)\subset O(5,2)$, i.e. by 
the energy level $E_0$ and a $USp(4)$ representation.

Let us analyse in details the case $N_f=6$. The singleton hypermultiplets
transform in the fundamental of $E_7$, which decomposes under $SO(12)\times 
O(2)$ as ${\bf 56}=({\bf 12},2)+({\bf 32},1)$. The hypermultiplets contain 
the fermion ($E_0=2$) and scalar ($E_0=3/2$) fields, $\psi^A, A^A_i$, where 
$A$ is an index in the ${\bf 56}$ of $E_7$ and $i$ is a index in the ${\bf 2}$ 
of $SU(2)$. The scalars obey the reality condition: 
$(A^{A}_i)^*=\epsilon_{ij}\Omega_{AB}A_j^B$ (where $\Omega_{AB}$ is the 
$E_7$ antisymmetric tensor). The total number of states is $4\times 56$.
The following bilinears belong to the energy-momentum tensor supermultiplet,
which contains $2^6$ states and is related to the graviton supermultiplet in 
$AdS_6$,
\be
{\rm graviton}\, (E_0=5):\, \, \, &\mbox{conserved traceless energy-momentum 
tensor} \nonumber\\ 
{\rm gravitinos}\, (E_0=9/2):\, \, \, &\gamma\partial A^A_i\gamma_\mu 
\psi^B\Omega_{AB}, \nonumber \\
SU(2)_R\, {\rm currents}\, (E_0=4):\, \,  \, 
&A^A_i\stackrel{\leftrightarrow}{\partial}_\mu A^B_j \Omega_{AB},\nonumber \\
{\rm antisymmetric\, tensor\, and \, singlet \, vector}\, (E_0=4):\, \, \, 
&\psi^A\sigma_{\mu\nu}\psi^B\Omega_{AB},\nonumber \\
{\rm fermions}\, (E_0=7/2):\, \, \, &A^A_i\psi^B\Omega_{AB},\nonumber \\
{\rm singlet\, scalar}\, (E_0=3):\, \, \, &A^A_iA^B_j\epsilon_{ij}\Omega_{AB}.
\ee
 The global current
supermultiplet, related to the $AdS_6$ vector multiplets, with $2^4\times 
\mbox{dim}G$ states, contains
\be
E_7 \, {\rm currents}\, (E_0=4):\, \, \, 
&A^A_i\stackrel{\leftrightarrow}{\partial}_\mu A^B_j\epsilon_{ij}T_{AB}^I + \psi^A\gamma_\mu\psi^B T_{AB}^I,\nonumber \\
{\rm fermions}\, (E_0=7/2):\, \, \, &A^A_i\psi^B T_{AB}^I,\nonumber \\
{\rm scalars}\, (E_0=3):\, \, \,   &\qquad\qquad\qquad\qquad\qquad A^A_iA^B_j\sigma_{ij}^a T_{AB}^I,\qquad (\mbox{dim}SU(2), \mbox{dim}E_7),\nonumber \\
(E_0=4):\, \, \,  &\qquad\qquad\qquad\qquad\qquad\psi^A\psi^B T_{AB}^I,\qquad\qquad\qquad (1,\mbox{dim}E_7),
\ee
where $T_{AB}^I$ are the $E_7$ (symmetric) matrix generators in the fundamental representation. Note that the total number of scalars is $4\, \mbox{dim}G+1$, in agreement with (\ref{manifold}).

The last set of scalars, singlets under $SU(2)_R$ and in the adjoint of $E_7$, are the highest components (highest number of $\theta$ and, as a consequence, also highest conformal dimension) of the global current supermultiplet and therefore correspond
to supersymmetry preserving deformations of the superconformal point. Having
conformal dimension 4, they correspond to relevant deformations, which break
superconformal invariance (and also the global symmetry $E_7$). Going to the 
Cartan subalgebra of $E_7$, we find $N_f$ parameters
$t_i$ ($i=0,...,N_f-1$). $t_0=1/g^2$ corresponds to turning on the inverse
coupling constant, breaking the global symmetry to $SO(2N_f)\times U(1)_I$,
which is appropriate for the Yang-Mills theory with non-infinite coupling. 
The other parameters $m_i$ corresponds to masses for some of the quarks and 
also break, partially or totally, $SO(2N_f)$. The other scalars in the global 
current multiplet, having lower dimension and therefore not being the highest 
components of their supermultiplet, are supersymmetry breaking deformations.

We see that also the scalar in the supergraviton multiplet, which is a 
singlet of
the R and global symmetries, is not the highest component of its 
supermultiplet and therefore it is a deformation which breaks supersymmetry. 
In particular, the
{\it dilaton} does not correspond to the coupling constant of the Yang-Mills 
theory. It is interesting to consider what happens to this singlet scalar in 
superconformal theories
in different dimensions. In $D=4$, it is the highest component in its 
multiplet (which is the graviton multiplet for $N=4$, a tensor multiplet for 
$N=2$ and
an hypermultiplet for $N=1$ \cite{FFZ}). In all the cases, it has dimension 
4. It corresponds to a marginal supersymmetric
deformation, which can be identified with the coupling constant: the theory
has indeed a line of fixed points. In $D=3$ and 6, the singlet is not the 
highest component of the supermultiplet and, therefore, it is a relevant deformation which breaks supersymmetry and it cannot be identified with the coupling constant. On the other hand, the (inverse) coupling constant is an irrelevant
deformation in $D=3$, which, therefore, does not belong to the massless multiplets in $AdS$, but to some massive KK mode. In $D=6$, the (inverse) coupling constant is a relevant parameter 
and belongs to some global current multiplet, as in the $D=5$ case.

Note that the five dimensional
Yang-Mills theories considered in this paper have been 
constructed  originally in terms of D4 and D8-branes systems. Also, we have 
argued an interpretation of their fixed 
point superconformal field theories (in the large $n$ limit) in terms of boundary singleton
theories of $F(4)$ supergravities in $AdS_6$. These two facts give evidence that a solitonic
D4-D8-branes configuration preserving eight supercharges, with a near
horizon geometry described by a $F(4)$ gauged six dimensional supergravity
theory (with 16 supercharges) should exist. In fact, since D8-branes are only
known to exist in the ``massive'' type IIA supergravity 
\cite{romans2, papado}, it may be natural to consider such solitonic solutions in
this ten dimensional supergravity. Such a relation between the $F(4)$ and the 
massive IIA supergravities is also  suggested by the fact that they 
are the only known cases where a Higgs mechanism takes place, where a
massless two-form absorbs the degrees a freedom of a gauge boson to
become massive. As a result, the relations between $p$-branes world-volumes
and $AdS_{p+2}$ theories that were already known for $p=3$ in type IIB 
and $p=2,5$ in eleven dimensional supergravity would be completed for
$p=4$ in the massive type IIA. 

Although the connection between $F(4)$ and massive type IIA supergravities
seems to suggest that the gauged six-dimensional supergravity, dual to the
five-dimensional fixed point, can be obtained as the near horizon geometry of
a configuration with D4 and D8 branes, it cannot be excluded that the brane configuration with $AdS_6$ near-horizon geometry is instead realized in a different set-up. A chain of dualities, for example, transforms the D4-D8 system into
a fivebrane wrapped around a circle in the $E_8\times E_8$ heterotic string.
In this heterotic description, the full series of $E_{N_f+1}$ global symmetries, which would be harder to get in a massive type IIA compactification, are more likely to be manifest as space-time fields, as discussed for six-dimensional
(1,0) theories with $E_8$ symmetry in \cite{berkooz}. The five-dimensional fixed points discussed in this paper are indeed the reduction to 5d (with additional Wilson line) of the six-dimensional (1,0) theory.

It would be interesting to find the right solution. Besides giving evidence for the $AdS$/CFT correspondence, it would provide an explicit KK reduction from ten or eleven dimensions to six, and the KK modes would give information on the spectrum of conformal operator of the fixed point theory.  

\noindent{\bf Acknowledgements} 

Research supported in part by
the EEC under TMR contract ERBFMRX-CT96-0090.
S.F. is supported in part by the DOE under grant DE-FG03-91ER40662, 
Task C, and
by ECC Science Program SCI$^*$ -CI92-0789 (INFN-Frascati).

\end{document}

\end{thebibliography}
\end{document}